%% file: main.tex
\documentclass[journal,10pt,twocolumn]{IEEEtran}
\usepackage{float}
\usepackage{amsmath,amsthm, amsfonts}

\input{notation}

\usepackage{algorithm}
\usepackage{algpseudocode}

\usepackage{graphicx}
\newtheorem{proposition}{Proposition}
\usepackage{array}
\usepackage{bm} 
\usepackage{booktabs}      % for \toprule, \midrule, \bottomrule
\usepackage{siunitx}       % for consistent units (optional but recommended)

\usepackage[most]{tcolorbox}
\tcbuselibrary{skins, breakable}

\usepackage{textcomp}
\usepackage{stfloats}
\usepackage{url}
\usepackage{verbatim}
\usepackage{tikz}

\usepackage[noadjust]{cite}
\usepackage{xcolor}
\usepackage{amsmath}
\allowdisplaybreaks[2]          % amsmath may break align at reasonable points

\usepackage{amssymb}
\usepackage{mathtools}
\usepackage{epstopdf}
\usepackage{graphicx}    % ensures \includegraphics is defined
\usepackage[font=small,skip=4pt,labelfont=bf, belowskip=10pt]{caption}

\usepackage{subcaption}

\usepackage{cuted}

\begin{document}
% \bstctlcite{BSTcontrol}
\title{{\huge A New Location Estimator for Mixed LOS \& NLOS scenarios}\\}

\author{
~Gaurav~Duggal,~R.~Michael~Buehrer,~Harpreet~S.~Dhillon, Jeffrey~H.~Reed
\thanks{\\
Gaurav Duggal, R. Michael Buehrer, Harpreet S. Dhillon, and Jeffrey H. Reed  are with Wireless@VT,  Bradley Department of Electrical and Computer Engineering, Virginia Tech,  Blacksburg,
VA, 24061, USA. Email: \{gduggal, rbuehrer, hdhillon, reedjh\}@vt.edu.\\ The support of NIST PSCR PSIAP through grant 70NANB22H070, NSF through grant CNS-1923807, CNS-2107276 and NIJ graduate research fellowship through grant 15PNIJ-23-GG-01949-RES is gratefully acknowledged.\\ }
}

% \markboth{Indoor locationing by Exploiting Diffraction}%


\maketitle

\begin{abstract}
Time-of-arrival (TOA)-based localization in mixed line-of-sight (LOS) and non-line-of-sight (NLOS) environments is challenging because conventional Euclidean range models do not capture diffraction-dominated propagation. We show that the diffraction path-length model smoothly transitions between LOS and diffraction-dominated NLOS conditions, eliminating the need for explicit path classification. Although this model provides a unified geometric description of mixed LOS/NLOS propagation, the resulting 3D maximum-likelihood problem is nonconvex, and a direct Gauss--Newton estimator based on this model can converge to suboptimal local minima. This motivates the development of a class of structure-exploiting estimators. For known target height, the model induces a virtual-anchor representation of the reduced 2D problem, enabling estimators that exhibit a clear complexity--performance tradeoff: surrogate formulations provide structure and computational efficiency, while a semidefinite-relaxation formulation more faithfully preserves the original likelihood at higher cost. Building on this same structure, we develop 3D sample--polish--select estimators that reduce the global search to one dimension, solve the associated fixed-height 2D subproblems, and then apply local nonlinear refinement in 3D. The proposed estimators achieve near-Cram\'er--Rao lower bound (CRLB) performance with substantially lower complexity than multistart Gauss--Newton, while also being far more robust to initialization than a direct single-start Gauss--Newton estimator.
\end{abstract}


\section{Introduction}
Time-of-arrival (TOA)-based wireless localization becomes substantially more challenging in non-line-of-sight (NLOS) environments. When the direct path between a transmitter and a receiver is blocked, the received signal may arrive through reflection, transmission, scattering, or diffraction, producing multiple paths with different delays. As a result, the measured TOA no longer corresponds directly to the true geometric distance, and localization methods based on a direct-path assumption can suffer severe performance degradation \cite{zekavat2019handbook}. This issue is particularly important in public-safety applications, where responders must be localized reliably inside buildings \cite{duggal_los_2023}. In such settings, even moderate ranging errors can lead to large position errors, potentially placing a responder in the wrong room or on the wrong floor.

Early work on TOA-based localization in mixed LOS/NLOS environments largely viewed NLOS measurements as harmful and therefore sought to identify and discard them prior to location estimation \cite{Wylie1996NLOS,Borras1998Decision,Chan2006TOA}. This viewpoint has remained prevalent in much of the subsequent literature \cite{li2026uwb}. 

A major conceptual shift was introduced by Qi \emph{et al.} \cite{Qi2002Prior,Qi2006Analysis,Qi2006TOA}, who proposed modeling NLOS measurements as the Euclidean range corrupted by a positive excess-delay term, commonly referred to as the NLOS bias, instead of treating such measurements as entirely unusable. Subsequent analyses by Jourdan \emph{et al.} \cite{Jourdan2008PEB} and Shen \emph{et al.} \cite{Shen2010Fundamental} established that NLOS measurements can still convey useful localization information when prior statistical knowledge of the bias is available, implying that outright rejection may also discard valuable locational information. Along similar lines, O’Lone \emph{et al.} \cite{OLone2019ExpBias,OLone2022TOAAOA} showed that, under a stochastic-geometry model with first-order (single-bounce) reflections, the bias of the first-arriving NLOS path admits a form that is well approximated by an exponential distribution, thereby providing analytical support for the exponential NLOS-bias models commonly adopted in the localization literature. Motivated by this perspective, a substantial body of work has focused on NLOS-bias mitigation \cite{guvenc2009survey}, including Bayesian/statistical approaches as well as optimization-based methods such as those by Venkatesh \emph{et al.} \cite{Venkatesh2007LP}, Vaghefi \emph{et al.} \cite{Vaghefi2012Cooperative,Vaghefi2013TOA}, and Chen \emph{et al.} \cite{Chen2012SDP}. Another line of work adopts a set theoretic approach for NLOS mitigation by Jia \emph{et al.}\cite{Jia2010Collaborative} and Dureppagari \emph{et al.} \cite{two_stage_ippm_Dureppagari}. 
\par
An alternate view of wireless localization in NLOS environments incorporates explicit awareness of the underlying signal propagation mechanism and is grounded in high-frequency ray-based electromagnetic modeling. In this viewpoint, Geometrical Optics (GO) arises as a short-wavelength asymptotic approximation to Maxwell's equations, under which signal propagation is represented by rays that travel in straight lines in homogeneous media and obey simple geometrical laws at smooth interfaces \cite{raytracing_yun}. This framework provides a natural basis for modeling NLOS propagation mechanisms such as specular reflection and wall transmission, while diffraction requires extensions beyond conventional GO, such as the Geometrical Theory of Diffraction (GTD) and the Uniform Theory of Diffraction (UTD).

In particular, Meissner \emph{et al.} \cite{Meissner2010VA} and subsequent works by Leitinger \emph{et al.} and Gentner \emph{et al.} \cite{Leitinger2014ML,Gentner2016ChannelSLAM} showed that, in NLOS localization scenarios, specular reflections can be modeled geometrically through the virtual-anchor concept, whereby certain reflected NLOS paths can be recast as LOS-like (Euclidean) constraints using building floor-plan information together with smooth-interface ray geometry governed by the law of reflection. This line of work further showed that not all multipath components are equally beneficial for localization: deterministic components, such as the direct path and resolvable specular reflections, carry useful location-related information, whereas diffuse multipath primarily degrades localization performance by reducing the quality of these informative components  \cite{Leitinger2015Info,Meissner2014Weighting,Witrisal2016FoeFriend}.
\par
Transmission through objects such as walls constitutes another possible NLOS propagation mechanism for wireless localization, particularly in Outdoor-to-Indoor (O2I) scenarios. Under a Euclidean approximation, the transmitted path is assumed to follow approximately the direct geometric path, with additional attenuation due to the material. This mechanism was considered in \cite{duggal2025}, where realistic EM-based ray-tracing simulations \cite{wirelessinsight} showed that wall transmission is negligible above 
$6$ GHz in the considered O2I scenario. This approximation assumes that the wall can be modeled as a single-layer, homogeneous, and symmetric dielectric slab \cite{yasmeen2023estimation}, and that its thickness is small relative to the overall propagation distance, so that refraction effects can be neglected.

\par
Edge diffraction has emerged as a key NLOS propagation mechanism in a range of settings, including O2I scenarios \cite{duggal20243d}, building corridors \cite{zhou2026novel}, urban environments \cite{tenerelli1998measurements}, and wireless sensing applications \cite{Pallaprolu}. The corresponding geometric path-length model was first introduced by Ang \emph{et al.} \cite{Ang1999DiffractionPoints}. Duggal \emph{et al.} later independently derived a mathematically equivalent formulation \cite{duggal2026optimalanchorplacementwireless} for localization and provided a more complete theoretical treatment by explicitly connecting it to Fermat’s principle of least time and the Keller cone associated with edge diffraction \cite{duggal20243d,duggal2025diffractionaidedwirelesspositioning}. In related localization work, Seow \emph{et al.} \cite{seow2008non} proposed a bidirectional TOA/AOA framework that exploits LOS and certain single-bounce NLOS paths; while not a dedicated diffraction-modeling study, their experiments included single-diffraction paths and treated higher-order diffraction as harmful multipath. By contrast, Duggal \emph{et al.} showed that localization can be achieved using only TOA measurements associated with diffraction paths, without requiring angle information, and demonstrated substantial gains over prior-art methods in diffraction-dominated scenarios, particularly in O2I environments \cite{duggal20243d,duggal2025diffractionaidedwirelesspositioning}. They further established the corresponding fundamental bounds for diffraction-aided localization and proposed a Gauss--Newton-based estimator, termed D-NLS, for the resulting nonlinear localization problem. However, in practice, D-NLS does not attain the CRLB on the localization accuracy in terms of Root Mean Squared Error (RMSE) for two main reasons. First, model mismatch arises when some ranging measurements don't correspond to the assumed diffraction path length model. Second, Being an iterative method, the estimator is sensitive to the initialization, whichcauses the Gauss--Newton iterations to converge to suboptimal local solutions. Developing a computationally efficient method to mitigate this initialization sensitivity is the primary focus of this paper. Our main contributions are:
\begin{itemize}
    \item \textbf{Unified LOS/NLOS path model and virtual-anchor formulation}: We show that the previously developed diffraction path model provides a unified geometric description for both LOS and diffraction-dominated NLOS propagation. The model transitions smoothly between these regimes as a function of the anchor--target geometry, thereby avoiding explicit LOS/NLOS labeling and separate treatment of the corresponding paths. Moreover, by exposing the mathematical structure of this unified path model, we derive a new \emph{virtual-anchor} construction, which serves as a key analytical tool and forms the foundation for the 2D and 3D localization methods developed in this paper.

    \item \textbf{2D localization algorithms as a building block for 3D localization}: For a fixed target height \(z\), we consider the reduced problem of estimating the horizontal target coordinates \((x,y)\).\footnote{If the vertical coordinate is known from auxiliary measurements, such as a barometer, or from \emph{a priori} information assuming a collaborative target, this reduced formulation also provides a standalone localization method.} This 2D formulation does not imply that the anchors and target are coplanar; rather, the anchors may remain arbitrarily distributed in 3D space, while only the target height is held fixed. Building on the virtual-anchor formulation, we develop two estimators based on a structured squared-range surrogate objective, including one that admits an exact solution, as well as a third estimator based directly on the exact maximum-likelihood objective via semidefinite relaxation. Together, these estimators reveal a fundamental complexity--performance tradeoff between tractable surrogate formulations and statistically faithful but less tractable exact formulations. These methods also serve as the core subproblem solvers in the proposed 3D framework, where \((x,y)\) is estimated for each candidate value of \(z\). Their performance is analyzed as standalone estimator as well by comparing with the CRLB.

    \item \textbf{3D localization algorithm}: Building on the preceding 2D framework, we develop a computationally efficient algorithm for estimating the full 3D target position \((x,y,z)\) under the unified LOS/NLOS path model. By leveraging the mathematical structure of the diffraction path model, we reformulate the nonconvex 3D maximum-likelihood estimation task as a one-dimensional search over the target height \(z\), while the corresponding horizontal coordinates are recovered by solving the 2D virtual-anchor-based subproblem developed in the previous contribution. This leads to a \emph{sample--polish--select} strategy that combines low-dimensional global exploration with local nonlinear refinement. In this way, the 2D methods serve as the key stepping stone that makes the 3D estimator computationally practical. Compared with conventional 3D iterative methods, the proposed approach offers significantly lower complexity and improved robustness to initialization, while achieving near-CRLB accuracy.
\end{itemize}

\section{System Model and Problem Formulation}
\begin{figure}[htbp]
    \centering
    \includegraphics[width=0.8\linewidth]{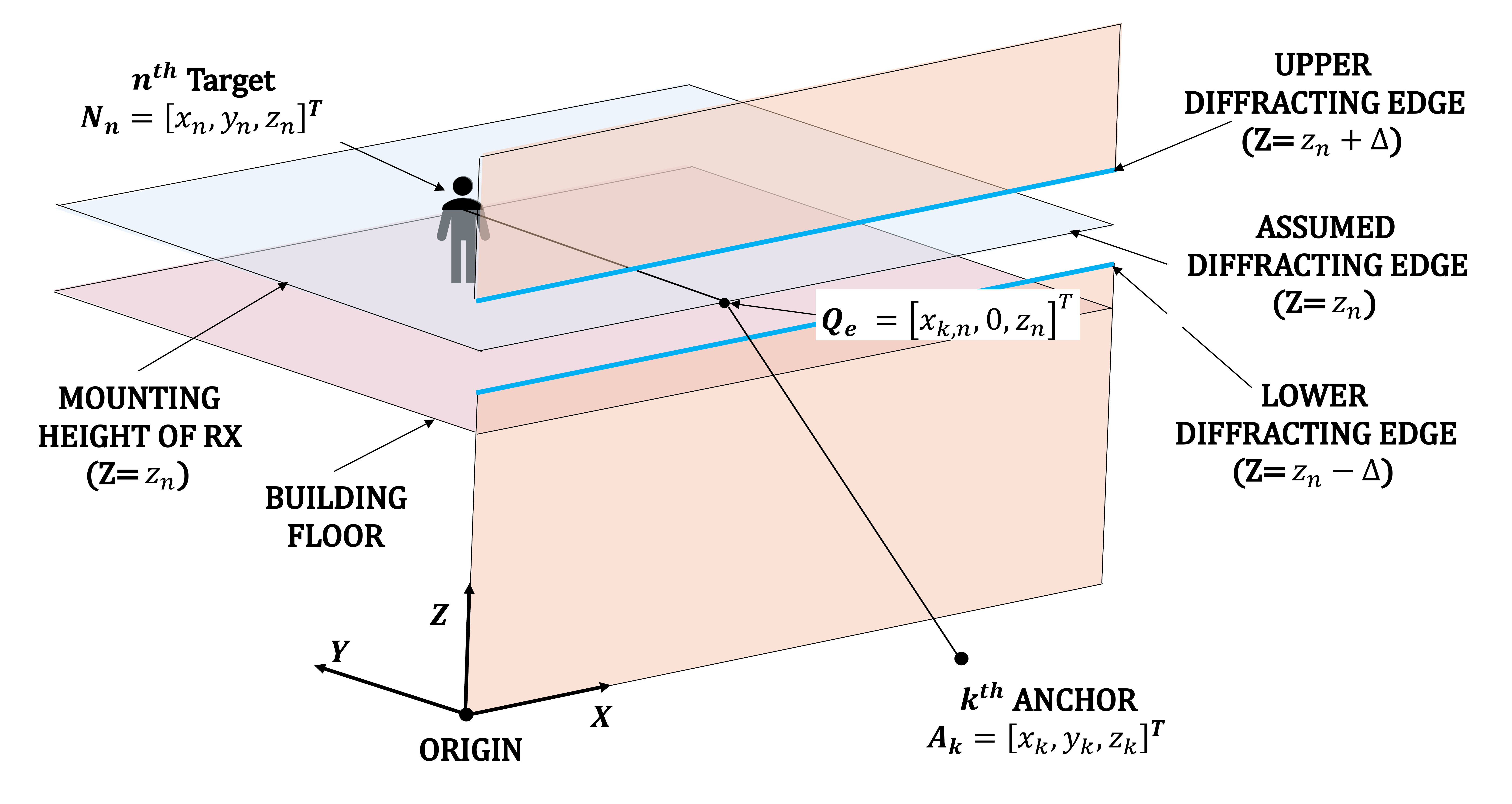}
    \caption{In the O2I scenario, we have \(K\) anchors transmitting orthogonal signals which are received by the $n^{\text{th}}$ target inside the building. 
The received signal includes several MPCs, from which the ranging measurement corresponding to the diffraction path length 
\(\bm{A_k}\bm{Q}_e\bm{N}_n\) is extracted. 
A bandwidth of \(200\,\text{MHz}\) is assumed, ensuring that all MPCs are resolvable.}

    \label{fig_system_model}
\end{figure}
\subsection{Geometric Setup}
We consider a mixed outdoor-to-indoor localization scenario with $K$ anchors deployed outside a building and a target located indoors. For each anchor $k\in\{1,\dots,K\}$, a scalar range measurement $r_k$ is available. We model $\{r_k\}$ as noisy observations of the propagation path length associated with edge-diffracted signal components (e.g., diffraction from window or building edges) \cite{duggal2025diffractionaidedwirelesspositioning}. We term these components \emph{unified LOS/NLOS} paths because the corresponding geometric path-length model varies continuously as the target transitions between line-of-sight (LOS) and non-line-of-sight (NLOS) configurations, thereby avoiding explicit LOS/NLOS labeling.

\subsection{Unified LOS/NLOS Path-Length Model}
\label{section_unified_path_length_model}
Let the $k^{th}$ anchor be located at $\bm{A}_k\triangleq[x_k,y_k,z_k]^T$ and the target at unknown coordinates $\bm{X}_{3\mathrm{D}}\triangleq[x,y,z]^T$. The unified path-length model provides a single geometric description that (i) coincides with the Euclidean distance in LOS and (ii) captures the additional detour induced by edge diffraction in NLOS. As the geometry varies, the model transitions smoothly between these regimes, so that the path classification is implicitly determined by the relative placement of the anchor, target, and diffracting edge. Therefore by using this path length model eliminates the need to determine if the path is LOS or NLOS. Following \cite{duggal2025diffractionaidedwirelesspositioning}, mathematically the path length is modeled as
\begin{align}
\label{eq_diffraction_path_length_approx}
p_k(\bm{X_{\text{3D}}})
= \sqrt{(x_k-x)^2 + \Big(\sqrt{y_k^2+(z_k-z)^2}+y\Big)^2}.
\end{align}

\begin{remark}[LOS/NLOS boundary condition]
When the anchor and target share the same horizontal plane (i.e., $z_k=z$), \eqref{eq_diffraction_path_length_approx} reduces to
\[
p_k(\bm{X_{\text{3D}}})=\sqrt{(x_k-x)^2 + (|y_k|+y)^2}.
\]
In this case, the diffraction-induced out-of-plane term vanishes and the model coincides with the standard Euclidean distance in the plane $Z=z$, recovering the LOS path-length model as a special case.
\end{remark}

% Equation
% \eqref{eq:pk_2d_diffraction} will serve as the basis for our 2D estimators.

% For notational convenience, we will sometimes collect the unknown horizontal
% coordinates into a vector
% \begin{equation}
%     \bm{X}_{\text{2D}} = [x,y]^\top \in \mathbb{R}^2,
% \end{equation}
% and view $p_k(x,y;z_0)$ as a function of $\bm{X}_{\text{2D}}$ and the known
% parameters $(\bm{A}_k, z_0)$.

\subsection{Range Measurement Model and R-LS (ML) Objective}
\label{subsec:measurement_model}

Each anchor $k$ obtains a noisy estimate of its propagation path length to the
target. We denote the range measurement at anchor $k$ by $r_k$, and assume a
standard additive noise model:
\begin{equation}
\label{eq:range_model_3d}
    r_k = p_k(\bm{X_{\text{3D}}}) + n_k,
    \quad k = 1,\dots,K,
\end{equation}
where $n_k$ represents the measurement error. We assume that
\begin{equation}
    n_k \sim \mathcal{N}(0,\sigma_{k}^2), \quad
    \text{independent across } k,
    \label{eq:noise_statistics}
\end{equation}
with known noise variances $\sigma_{k}^2$. The 3D range observation vector is
\begin{equation}
    \bm{r} = [r_1,\dots,r_{K}]^T \in \mathbb{R}^{K}.
\end{equation}

Under the Gaussian noise assumption, the negative log-likelihood is
\begin{equation}
    J_{\text{R-LS}}(\bm{X_{\text{3D}}})
    \propto \sum_{k=1}^{K} 
      w_k\bigl(p_k(\bm{X_{\text{3D}}}) - r_k\bigr)^2.
    \label{eq:J_ML_3d}
\end{equation}
Minimizing $J_{\text{R-LS}}(\bm{X_{\text{3D}}})$ with respect to $\bm{X_{\text{3D}}}=[x,y,z]^T$ therefore yields
the maximum-likelihood (ML) estimate
\begin{equation}
   \bm{X}_{\text{3D}}^* = \arg\min_{\bm{X}_{\text{3D}}\in\mathbb{R}^3} J_{\text{R-LS}}(\bm{X}_{\text{3D}}).
    \label{eq:est_RLS_3d}
\end{equation}

In the reduced 2D model (with $z=z_0$ fixed), the unknown target location is
$\bm{X}_{\text{2D}}=[x,y]^T$. The path length term in \eqref{eq:range_model_3d}
is replaced by \eqref{eq:pk_2d_diffraction}, yielding the ML objective
\begin{equation}
    J_{\text{R-LS}}(\bm{X}_{\text{2D}};z_0)
    \propto \sum_{k=1}^{K} w_k\,\bigl(p_k(\bm{X}_{\text{2D}};z_0) - r_k\bigr)^2,
    \quad
    w_k \triangleq \frac{1}{\sigma_{k}^2}.
    \label{eq:J_RLS_2d}
\end{equation}

We refer to \eqref{eq:J_RLS_2d} as the \emph{Range-domain Least-Squares
(R-LS)} cost, and its minimizer
\begin{equation}
    \arg\min_{\bm{X_{\text{2D}}}\in\mathbb{R}^2} J_{\text{R-LS}}(\bm{X_{\text{2D}}};z_0),
    \label{eq:est_RLS_2d}
\end{equation}
as the R-LS estimator for the 2D unified LOS/NLOS model.

The optimization problems \eqref{eq:est_RLS_3d} and \eqref{eq:est_RLS_2d} are
nonconvex with respect to the unknown coordinates $\bm{X_{\text{3D}}}$ and $\bm{X_{\text{2D}}}$. This is due to the nonlinear dependence of $p_k(\cdot)$ on the
unknown coordinates. As a result, naive gradient-based methods may converge to local minima or stationary points that are far from the global optimum. 
% to obtain more tractable formulations based on \emph{virtual anchors} and
% \emph{squared-range least-squares} (SR-LS), and we connect these to the
% framework of Beck and Stoica for range-based source localization \cite{beck2008exact}.

\section{Virtual Anchors and connection to Euclidean Ranging}
In this section, we exploit the structure of the proposed path-length model to develop the localization methods used in the rest of the paper. The resulting 3D estimation problem is nonconvex because the path-length expressions in \eqref{eq_diffraction_path_length_approx} contain nested square-root terms, making direct maximum-likelihood (ML) optimization challenging and motivating the Gauss--Newton approach used in prior work. To address this, we decompose the 3D problem into a family of 2D localization subproblems parameterized by the target height \(z\). For each fixed \(z\), we solve the corresponding 2D problem in \((x,y)\), and then perform a one-dimensional search over \(z\). As shown later, this provides an efficient route to 3D localization and also serves as a standalone 2D localization framework when \(z\) is known from \emph{a priori} information or auxiliary sensing. We begin with a {\em virtual-anchor} formulation for the fixed-\(z\) subproblem, which underpins all subsequent algorithmic developments. We then develop two approaches for this subproblem: a surrogate method based on a statistical approximation of the ML objective, and a semidefinite-relaxation method based on an optimization approximation of the lifted ML formulation. These two approaches yield different performance and computational complexity tradeoffs, which are examined later in the paper.

\subsection{Virtual-Anchor Embedding}
\label{section_reduced_2d_model}

We now exploit a key structural property of the proposed path-length model in \eqref{eq_diffraction_path_length_approx}. In particular, the model is symmetric in the horizontal coordinates \((x,y)\) but asymmetric in the vertical coordinate \(z\). This motivates treating the target height $z=z_0$ as fixed or known and reformulating the localization problem as the estimation of the target's horizontal coordinates \(\bm{X}_{\text{2D}}=[x,y]^T\) from the available ranging measurements. Under this assumption, the unified LOS/NLOS path from anchor \(k\) to the target admits a simplified form. Let
\begin{equation}
    r_{\perp,k}(z_0)
    \triangleq \sqrt{y_k^2 + (z_k - z_0)^2}
    \label{eq:r_perp_def}
\end{equation}
denote a \emph{perpendicular distance} term that depends only on the anchor
location and the known target height \(z_0\). Intuitively, \(r_{\perp,k}\) captures
the effective vertical and lateral separation between anchor \(k\) and the target
in the plane orthogonal to the diffraction edge.

Using~\eqref{eq:r_perp_def}, the unified LOS/NLOS path length for fixed
\(z_0\) can be written as
\begin{equation}
    p_k(\bm{X}_{\text{2D}}; z_0)
    = \sqrt{(x - x_k)^2 + \bigl(y + r_{\perp,k}(z_0)\bigr)^2}
    \label{eq:pk_2d_diffraction}
\end{equation}

Now, define the location of the virtual anchors as
\begin{align}
\label{eq_virtual_anchor_location}
\bm{\tilde{a}}_k \triangleq
\begin{bmatrix}
x_k\\
-\,r_{\perp,k}(z_0)
\end{bmatrix}
\in \mathbb{R}^2,
\qquad
k=1,\ldots,K.
\end{align}
This leads to the following observation.

\begin{proposition}[Virtual-anchor embedding]
\label{prolocation:1_virtual_anchor}
For a fixed target height \(z_0\), the 2D reduced path model in \eqref{eq:pk_2d_diffraction} is exactly equivalent to a Euclidean ranging model in a virtual plane:
\begin{align}
p_k(\bm{X}_{\text{2D}};z_0) = \lVert\bm{X}_{\text{2D}}-\bm{\tilde{a}}_k\rVert_2,\qquad k=1,\ldots,K,
\label{eq:virtual_anchor_euclidean}
\end{align}
\end{proposition}

Thus, by fixing \(z_0\), the original mixed LOS/NLOS path model admits an exact Euclidean embedding in terms of virtual anchors. This embedding will serve as the basis for simplifying the resulting 2D localization problem and for developing the estimators in the subsequent subsections.

\subsection{SR-LS formulations}
\label{section:SR_LS_formulations}
In this section we aim to replace the R-LS (ML) objective in \eqref{eq:J_RLS_2d} with an approximate but more tractable surrogate. We begin by squaring the measurement model in \eqref{eq:range_model_3d}. For notational simplicity we note the target's 3D coordinates can be expressed as $X_{\text{3D}}=[X_{\text{2D}},z_0]^T$ hence we replace $p_k(X_{\text{3D}})$ by $p_k(X_{\text{2D}},z_0)$ and since $z_0$ is assumed fixed, we drop it from our notation to obtain
\begin{equation}
\begin{aligned}
r_k^2
&= \bigl(p_k(\bm{X_{\text{2D}}}) + n_k\bigr)^2 \\
&= (p_k(\bm{X_{\text{2D}}}))^2 + 2\,p_k(\bm{X_{\text{2D}}})\,n_k + n_k^2. 
\label{eq:squared_range_measurement_model}
\end{aligned} 
\end{equation}

Define 
\[
\epsilon_k(\bm{X_{\text{2D}}}) \triangleq 2\,p_k(\bm{X_{\text{2D}}})\,n_k + n_k^2. 
\]
\textbf{Approximation 1}: If the range noise variance is small relative to the path length (typical at moderate-to-high SNR) then $n_k^2$ can be ignored and we approximate
\[
\epsilon_k(\bm{X_{\text{2D}}}) \approx 2p_k(\bm{X_{\text{2D}}})n_k.
\]
Under the zero-mean assumption on the range noise $n_k$ in \eqref{eq:noise_statistics}, the induced error term remains zero-mean; however, its variance scales with the unknown path length $p_k(\bm{X}_{\text{2D}})$. Thus, squaring the observation model introduces \emph{heteroscedastic} noise.
\begin{equation}
\mathrm{Var}\!\left(\varepsilon_k(\bm{X}_{\text{2D}})\right) \approx 4\,p_k^2(\bm{X}_{\text{2D}})\,\sigma_k^2.
\label{eq:epsilon_variance}
\end{equation}
\textbf{Approximation 2}: Ignore the dependence of the noise variance on $X_{\text{2D}}$.
A full Gaussian maximum-likelihood formulation for \eqref{eq:squared_range_measurement_model} would weight each term by
the inverse of \eqref{eq:epsilon_variance}, which depends on the unknown $\bm{X}_{\text{2D}}$ through
$p_k(\bm{X}_{\text{2D}})$. To obtain a tractable surrogate, we approximate this dependence by using fixed
(precomputable) weights $w_k$ (e.g., via a proxy for $p_k(\bm{X}_{\text{2D}})$ such as $r_k$), yielding the squared-range
least-squares objective (SR-LS) as
\begin{equation}
J_{\text{SR-LS}}(\bm{X}_{\text{2D}})
\triangleq
\sum_{k=1}^{K} \Tilde{w}_k\Bigl(r_k^2 - p_k^2(\bm{X}_{\text{2D}})\Bigr)^2,\; \Tilde{w}_k = \frac{1}{4\sigma_k^2r_k^2} 
\label{eq:J_SRLS_2d}
\end{equation}
Thus, under these two approximations, the SR-LS objective can be viewed as an approximate maximum-likelihood criterion for the original problem and hence serves an approximation to the R-LS objective. Next, we show that the SR-LS objective is simpler to handle as an optimization problem. We begin by defining an augmented variable $\bm{u}\in \mathbb{R}^3$ as below. 
\begin{align}
\bm{u} \triangleq\begin{bmatrix}
    \bm{X_{\text{2D}}} \\
    \left\lVert \bm{X}_{\text{2D}} \right\rVert_2^2 
    \end{bmatrix},\; \left\lVert \bm{X}_{\text{2D}} \right\rVert_2^2 = x^2+y^2.
    \label{eq:lifted_optimization_variable}
\end{align}
Using this we can rewrite the square of the unified LOS/NLOS path in \eqref{eq:virtual_anchor_euclidean} as an affine function of $\bm{u}$ as
\begin{align}
\label{eq:squared_path_length_affine_form}
\bigl(p_k(\bm{X}_{\text{2D}};z_0)\bigr)^2 = \bm{q_k}^T\bm{u}+c_k, \\
\bm{q_k} \triangleq \begin{bmatrix}
    -2x_k\\
    2r_{\perp,k}\\ \nonumber
    1
\end{bmatrix}, \; c_k \triangleq \lVert\bm{\tilde{a}_k}\rVert_2^2.
\end{align}

\subsubsection{Formulation 1: SR-LS GTRS formulation}
\label{section_GTRS_formulation}
Using the affine function representation of the squared path length from \eqref{eq:squared_path_length_affine_form}, for each anchor $k$ in \eqref{eq:J_SRLS_2d}, we define the SR-LS residual as
\[
e_k(\bm{u}) \triangleq r_k^2 - p^2_k(\bm{X}_{\text{2D}}) = r_k^2 - (\bm{q_k}^T\bm{u}+c_k).
\] 
Additionally define
\[
\delta_k \triangleq r_k^2-c_k.
\]
So the total weighted SR-LS cost is
\begin{equation}
\begin{aligned}
    J_{\text{SR-LS}}(\bm{X}_{\text{2D}}) &=
\sum_{k=1}^{K} \Tilde{w}_k(e_k(\bm{u}))^2  \\
&= \sum_{k=1}^{K}\Tilde{w}_k(-\bm{q_k}^T\bm{u}+\delta_k)^2  \\
&= \bm{u^T}\bm{M}\bm{u} + 2\bm{m^T}\bm{u}+\gamma.
\end{aligned}
\end{equation}

Here, the quadratic, linear and constant terms are
\begin{equation}
\begin{aligned}
    \bm{M}\triangleq \sum_{k=1}^{K} \Tilde{w}_k\bm{q_k}\bm{q_k^T},\;\bm{m}\triangleq \sum_{k=1}^{K} -\Tilde{w}_k\delta_k\bm{q_k}, \gamma \triangleq \sum_{k=1}^{K} \Tilde{w}_k\delta_k^2.
\end{aligned}
\label{eq:defn_M_m_delta}
\end{equation}

Thus we have expressed the SR-LS cost as a quadratic function of the lifted variable $\bm{u}$. From the definition of $\bm{M}$ in \eqref{eq:defn_M_m_delta}, it is easy to see that since it is the sum of the outer products of $\bm{q_k}$ and $w_k\geq 0$, we have $\bm{M}\succeq0$ hence the objective is convex.
Note that there is a single quadratic constraint which is obtained from the definition of our optimization variable $\bm{u}$ in \eqref{eq:lifted_optimization_variable}. Essentially we must have
\[
u_3 = u_1^2+u_2^2, 
\]
where $u_i$ represents the vector's $i^{\text{th}}$ entry.
This is a quadratic constraint in terms of $\bm{u}$ and we define the following matrices
\begin{align}
    \bm{H}\triangleq
    \begin{bmatrix}
        1&0&0 \\
        0&1&0 \\
        0&0&0 \\
    \end{bmatrix}, \; \bm{h} \triangleq \begin{bmatrix}
        0\\0\\-0.5 \nonumber
    \end{bmatrix}
\end{align}
to express it in matrix form. The complete optimization problem formulation is given as
\begin{equation}
\boxed{
\begin{aligned}
\min_{\bm{u}\in\mathbb{R}^3}\quad & f(\bm{u})=\bm{u^T}\mathbf{M}\bm{u} + 2\,\bm{m^T}\bm{u} + \gamma \\
\text{s.t.}\quad & g(\bm{u})=\bm{u^T}\mathbf{H}\bm{u} + 2\,\bm{h^T}\bm{u} = 0 .
\end{aligned}
}
\label{eq:2D_SR_LS_GTRS}
\end{equation}

Thus, by assuming $\bm{u}$ as our optimization variable, we can express our SR-LS cost as a convex quadratic function with a single nonconvex quadratic equality constraint. This is a least squares optimization problem that can be expressed as a Generalized Trust Region Subproblem (GTRS) as in \cite{beck2008exact}.
\par
\par
The GTRS in \eqref{eq:2D_SR_LS_GTRS} is solved by reducing its KKT conditions to a scalar root-finding problem in the Lagrange multiplier $\lambda$. Specifically, stationarity gives $\bm u(\lambda)$ as the solution of
$(\mathbf{M}+\lambda\mathbf{H})\bm u=-(\bm m+\lambda\bm h)$, while feasibility requires $\phi(\lambda)\triangleq \bm u(\lambda)^T\mathbf{H}\bm u(\lambda)+2\bm h^T\bm u(\lambda)=0$. We bracket the root over the interval where $\mathbf{M}+\lambda\mathbf{H}\succ 0$ and solve $\phi(\lambda)=0$ via bisection. For each trial $\lambda$, $\bm u(\lambda)$ is computed efficiently by Cholesky factorizing $\mathbf{M}+\lambda\mathbf{H}$ and solving the resulting triangular systems, thereby avoiding explicit matrix inversion \cite{golub2013matrix}. This yields $\bm u^\star=\bm u(\lambda^\star)$ and $\bm{X}_{\text{2D}}^{\text{GTRS}}=[u_1^\star,u_2^\star]^T$. The KKT reduction and monotonicity of $\phi(\lambda)$, which ensure uniqueness of the root and justify bisection, are detailed in Appendix~\ref{section_KKT}, while Algorithm~\ref{alg:gtrs_1d} summarizes the implementation. Under standard GTRS regularity conditions, this procedure is exact and returns the global optimizer of \eqref{eq:2D_SR_LS_GTRS}, with no relaxation gap.

\begin{algorithm}[t]
\caption{1D bisection solver for SR-LS GTRS in \eqref{eq:2D_SR_LS_GTRS}}
\label{alg:gtrs_1d}
\begin{algorithmic}[1]
\Require $\bm{M},\bm m,\bm{H},\bm h$ from \eqref{eq:defn_M_m_delta}; tolerance $\varepsilon$; max iters $N_{\max}$.
\Ensure $\bm u^\star$ and $\bm X_{\text{2D}}^{\text{GTRS}}=[u_1^\star,u_2^\star]^T$.
\Statex
\State \textbf{Define} $\bm u(\lambda)$ via $(\mathbf{M}+\lambda\mathbf{H})\bm u=-(\bm m+\lambda\bm h)$ and
$\phi(\lambda)\triangleq \bm u(\lambda)^T\mathbf{H}\bm u(\lambda)+2\bm h^T\bm u(\lambda)$.
\Statex

\State \textbf{(Bracket)} Find $\lambda_L,\lambda_U$ such that $\mathbf{M}+\lambda\mathbf{H}\succ 0$ on $[\lambda_L,\lambda_U]$ and $\phi(\lambda_L)\ge 0\ge \phi(\lambda_U)$.
\Comment{e.g., start from any PD $\lambda$ and expand until a sign change}

\For{$t=1$ \textbf{to} $N_{\max}$}
\State $\lambda\gets(\lambda_L+\lambda_U)/2$; compute $\phi(\lambda)$.
\If{$|\phi(\lambda)|\le \varepsilon$} \State \textbf{break} \EndIf
\If{$\phi(\lambda)>0$} \State $\lambda_L\gets\lambda$ \Else \State $\lambda_U\gets\lambda$ \EndIf
\EndFor

\State $\bm u^\star\gets \bm u(\lambda)$; \quad $\bm{ X}_{\text{2D}}^{\text{GTRS}}\gets [u_1^\star,u_2^\star]^T$; \Return
\end{algorithmic}
\end{algorithm}

\subsubsection{Formulation 2: Unconstrained Squared Ranges (USR) formulation}
\label{section_USR_formulation}
The GTRS formulation in \eqref{eq:2D_SR_LS_GTRS} enforces the quadratic consistency constraint that couples the lifted variables. If we \emph{drop the constraint in} \eqref{eq:2D_SR_LS_GTRS}, the problem reduces to a purely \emph{weighted linear least-squares (WLLS)} estimation of the lifted vector $\bm{u}$. Specifically, using the linear model $r_k^2-c_k=\bm{q}_k^T\bm{u}$, define the stacked measurement vector and design matrix as
\begin{equation}
\begin{aligned}
\bm{Q} \triangleq \begin{bmatrix}\bm{q}_1^T \\ \vdots \\ \bm{q}_K^T\end{bmatrix},\quad
\bm{b} \triangleq \begin{bmatrix}r_1^2-c_1 \\ \cdots \\ r_K^2-c_K\end{bmatrix}, 
\end{aligned}
\label{eq:Q_b}
\end{equation}
and the diagonal weight matrix $\bm{\Tilde{W}}\triangleq \mathrm{diag}(\Tilde{w}_1,\ldots,\Tilde{w}_K)$. The unconstrained SR-LS estimate is then obtained in one shot as
\begin{equation}
\boxed{
\begin{aligned}
\bm{u}^{*}
&= \arg\min_{\bm{u}\in\mathbb{R}^3}\ \left\|\bm{\Tilde{W}}^{1/2}\left(\bm{b}-\bm{Q}\bm{u}\right)\right\|_2^2 \\
&= \left(\bm{Q}^T\bm{\Tilde{W}}\bm{Q}\right)^{-1}\bm{Q}^T\bm{\Tilde{W}}\,\bm{b},
\end{aligned}
}
\label{eq:WLLS_solution}
\end{equation}
provided $\bm{Q}^T\bm{W}\bm{Q}$ is nonsingular. The corresponding location estimate is
\begin{equation}
\bm{X}_{\text{2D}}^{\text{USR}}=\begin{bmatrix}u^*_1 & u^*_2\end{bmatrix}^T.
\end{equation}    

\subsection{R-LS formulation}
We next consider a formulation that more directly addresses the maximum-likelihood (ML) problem by minimizing the range-domain least-squares (R-LS) objective in \eqref{eq:J_RLS_2d}, which fits the measured path lengths directly in the range domain. In contrast to the surrogate approach, this method preserves the original ML objective but handles its nonconvexity through an optimization relaxation.
\subsubsection{Formulation 3: SemiDefinite Relaxation (SDR)}
\label{section_SDR_formulation}
Specifically, for fixed target height \(z=z_0\), we rewrite the minimization problem in \eqref{eq:est_RLS_2d} in terms of the 2D target coordinates \(\bm{X}_{\text{2D}}\), introduce an augmented variable to lift the problem to a higher-dimensional space, and then apply a semidefinite relaxation (SDR). Substituting the
fixed-height path model via the virtual anchor
representation in \eqref{eq:virtual_anchor_euclidean}, the objective becomes
\begin{align}
\min_{\bm{X}_{\text{2D}}} \sum_{k=1}^{K} w_k
\Big(r_k-\|\bm{X}_{\text{2D}}-\bm{\tilde a_k}\|_2\Big)^2 .
\end{align}
\paragraph{Augmented optimization variable} To express this objective in a
form amenable to convex relaxation, we define an augmented
variable
\begin{align}
\bm{\bar v} \triangleq
\begin{bmatrix}
\bm{X}_{\text{2D}} \\
\bm{t} \\
1
\end{bmatrix}
=
\begin{bmatrix}
x \\
y \\
t_1 \\
\vdots \\
t_K \\
1
\end{bmatrix}
\in \mathbb{R}^{K+3},
\quad
\bm{t} \triangleq [t_1,\ldots,t_K]^T ,
\end{align}
where $t_k$ is an auxiliary variable which denotes the predicted range associated with the
$k^\text{th}$ virtual anchor, i.e., $t_k \approx \lVert\bm{X}_{\text{2D}}-\bm{\tilde a_k}\rVert_2$.
Next, define the lifted Gram matrix
\begin{align}
\bm{Z} \triangleq \bm{\bar v} \bm{\bar v}^T \in \mathbb{S}_+^{K+3}.
\end{align}
Note, the variables of interest - the target's 2D location, can be recovered directly from the last column of $\bm{Z}$ as
\begin{equation}
\begin{aligned}
x = \bm{Z}_{1,K+3}, \quad
y = \bm{Z}_{2,K+3}, \\
t_k = \bm{Z}_{k+2,K+3}, \quad
t_k^2 = \bm{Z}_{k+2,k+2},
\end{aligned}
\label{eq_define_optimization_variables_gram_matrix}
\end{equation}
where $\bm{A}_{i,j}$ denotes the $(i,j)$th entry of the matrix $\bm{A}$.
\paragraph{Linearizing the R-LS objective and distance constraints}
Using \eqref{eq_define_optimization_variables_gram_matrix}, each term in the R-LS objective can be written as
\[
(r_k-t_k)^2 = r_k^2 - 2r_k \bm{Z}_{k+2,K+3} + \bm{Z}_{k+2,k+2}.
\]
Hence the full objective becomes linear in the matrix
$\bm{Z}$:
\begin{equation}
\begin{aligned}
\sum_{k=1}^{K} w_k (r_k-t_k)^2
&= \sum_{k=1}^{K} w_k \Bigl(
   r_k^2 - 2r_k \bm{Z}_{k+2,K+3} \\
&\qquad\qquad\quad
   + \bm{Z}_{k+2,k+2}
   \Bigr).
\end{aligned}
\label{eq_SDP_objective}
\end{equation}
Next, using the virtual-anchor representation in \eqref{eq:virtual_anchor_euclidean}, we have
\begin{align}
t_k^2
= \|\bm{X}_{\text{2D}}-\bm{\tilde a}_k\|_2^2
= x^2+y^2-2\bm{\tilde a}_k^T \bm{X}_{\text{2D}}+\|\bm{\tilde a}_k\|_2^2.
\label{eq_auxiliary_var_t_los_nlos_path_relation}
\end{align}
Substituting the lifted variables from \eqref{eq_define_optimization_variables_gram_matrix} into \eqref{eq_auxiliary_var_t_los_nlos_path_relation}, we obtain
the affine equality
\begin{align}
\bm{Z}_{k+2,k+2}
=
\bm{Z}_{1,1}+\bm{Z}_{2,2}
-2\bm{\tilde a}_k^T
\begin{bmatrix}
\bm{Z}_{1,K+3}\\
\bm{Z}_{2,K+3}
\end{bmatrix}
+\|\bm{\tilde a}_k\|_2^2.
\label{eq_sdp_contraint_1}
\end{align}
Thus, the target location variables $(x,y)$ and the range variables $\{t_k\}$ are coupled
directly within a single positive semidefinite matrix.
\paragraph{Triangle inequality tightening} For any two virtual
anchors, indexed by $k_1,k_2$, we define the distance between them as
\[
d_{k_1,k_2} \triangleq \|\bm{\tilde a}_{k_1}-\bm{\tilde a}_{k_2}\|_2 .
\]
For a target located at $X_{2D}$, the triangle inequality
gives
\[
|t_{k_1}-t_{k_2}| \le d_{k_1,k_2},
\]
which is equivalent to
\[
(t_{k_1}-t_{k_2})^2 \le d_{k_1,k_2}^2.
\]
Using the lifted matrix $\bm{Z}$, this becomes
\begin{align}
\bm{Z}_{k_1+2,k_1+2}
+
\bm{Z}_{k_2+2,k_2+2}
-
2\bm{Z}_{k_1+2,k_2+2}
\le d_{k_1,k_2}^2.
\label{eq_sdp_contraint_2}
\end{align}

In addition, the triangle inequality implies
\[
t_{k_1}+t_{k_2} \ge d_{k_1,k_2},
\]
which can be written as the linear constraint
\begin{align}
\bm{Z}_{k_1+2,K+3} + \bm{Z}_{k_2+2,K+3} \ge d_{k_1,k_2}.
\label{eq_sdp_contraint_3}
\end{align}
\paragraph{Final RLS objective Semidefinite Program (SDP)}
\label{section:R_LS_formulation}
The objective in \eqref{eq_SDP_objective} is expressed as a linear function of the optimization variable $\bm{Z}$. Furthermore, the constraints in \eqref{eq_sdp_contraint_1}--\eqref{eq_sdp_contraint_3} are all affine in $\bm{Z}$. In the exact lifted formulation, the Gram matrix corresponding to the augmented variable matrix is required to be positive semidefinite and rank one, and the rank-one condition introduces nonconvexity. Relaxing this condition yields a convex semidefinite program, given by

\begin{equation}
\boxed{
\begin{aligned}
\min_{\bm{Z}} \;
& \sum_{k=1}^{K} w_k \bigl(
r_k^2 - 2r_k \bm{Z}_{k+2,K+3}
+ \bm{Z}_{k+2,k+2}
\bigr)\\
\text{s.t.}\;
& \bm{Z} \succeq 0,\quad \bm{Z}_{K+3,K+3}=1,\bm{Z}_{k+2,K+3}\ge 0,\\
& \bm{Z}_{k+2,k+2}
= \bm{Z}_{1,1}+\bm{Z}_{2,2}
-2\bm{\tilde a}_k^T
\begin{bmatrix}
\bm{Z}_{1,K+3}\\
\bm{Z}_{2,K+3}
\end{bmatrix}
+\|\bm{\tilde a}_k\|_2^2,
\\
&\qquad\qquad\quad\forall k \in \mathcal{K},\\
& \bm{Z}_{k_1+2,k_1+2}
+\bm{Z}_{k_2+2,k_2+2}
-2\bm{Z}_{k_1+2,k_2+2}
\le d_{k_1,k_2}^2,
\\
& \bm{Z}_{k_1+2,K+3}
+\bm{Z}_{k_2+2,K+3}
\ge d_{k_1,k_2},\\
&\qquad\qquad\quad \forall k_1,k_2\in \mathcal{K}, k_1 \neq k_2.\\
\end{aligned}
}
\end{equation}

where 
\[
\mathcal{K} \triangleq \{1,2,\cdots,K\}
\]
represents the anchor indices.
The relaxed SDP can be solved using standard commercial solvers to obtain the optimal matrix $\bm{Z}^\star$. The corresponding 2D target location estimate is then recovered directly from $\bm{Z}^\star$ as
\begin{align}
\bm{X}^{\mathrm{SDR}}_{2D}
=
\begin{bmatrix}
\bm{Z}^\star_{1,K+3}\\
\bm{Z}^\star_{2,K+3}
\end{bmatrix}.
\end{align}

\section{3D location estimation}
\label{section_3d_location_estimation}
In this section, we develop the proposed 3D localization method by leveraging the structure revealed by the preceding 2D formulation. For a fixed target height \(z\), the corresponding horizontal localization problem in \((x,y)\) can be solved to reduce the original 3D range-domain least-squares (R-LS) objective to a scalar function of \(z\), which we refer to as the \(z\)-profile. This profiling viewpoint transforms the original nonconvex 3D estimation problem into a one-dimensional search over the parameter $z$. In practice, however, the 2D subproblem is solved using the approximate methods developed in Section~III, namely GTRS, USR, and SDR, so the resulting \(z\)-profile is also approximate. Motivated by this, we use the approximate profile only to generate promising 3D seeds, which are then refined using Gauss--Newton iterations. This leads to a sample-polish-select procedure: candidate heights are first sampled, each sample is mapped to a 3D seed through the corresponding 2D solution, the resulting seeds are polished by Gauss--Newton iterations, and the final estimate is chosen as the one attaining the smallest R-LS cost.
\subsection{Profiling out the horizontal coordinates}
We can express the 3D target coordinates as
\begin{equation}
    \bm{X}_{\text{3D}} \triangleq \begin{bmatrix}
    \bm{X}_{\text{2D}} \\
    z
    \end{bmatrix}.
\end{equation}
For any fixed height $z$, we define the inner 2D subproblem as
\begin{equation}
\label{eq:true_inner_problem}
    \bm{X}_{\text{2D}}^*(z) = \arg \min_{\bm{X_{\text{2D}}}\in \mathbb{R}^2} J_{\text{R-LS}}\Bigl(\begin{bmatrix}
    \bm{X}_{\text{2D}} \\
    z
    \end{bmatrix}\Bigr)
\end{equation}
where $J_{\text{R-LS}}(\bm{X}_{\text{3D}})$ is the objective function defined in \eqref{eq:J_ML_3d} and the superscript $*$ represents the optimal solution (optimizer) of the minimization problem. Next the associated z-profiled objective is defined as
\begin{equation}
\begin{aligned}
J_{\text{prof.}}(z) &\triangleq \min_{\bm{X_{\text{2D}}}\in \mathbb{R}^2} J_{\text{R-LS}}\Bigl(\begin{bmatrix}
    \bm{X}_{\text{2D}} \\
    z
    \end{bmatrix}\Bigr)\\
    &= J_{\text{R-LS}}\Bigl(\begin{bmatrix}
    \bm{X}_{\text{2D}}^*(z) \\
    z
    \end{bmatrix}\Bigr).
\end{aligned}
\label{eq:z_profile_objective}
\end{equation}

Note $J_{\text{prof.}}(z)$ is the best achievable 3D residual when the height is $z$, i.e. after optimally choosing $\bm{X}_{\text{2D}}$.
\par
In the ideal case where the inner 2D subproblem is solved globally for each $z$, profiling preserves the global optimum value
\begin{equation}
    \min_{z\in \mathbb{R}} J_{\text{prof.}}(z) = \min_{\bm{X}_{\text{3D}}\in \mathbb{R}^3} J_{\text{R-LS}}(\bm{X}_{\text{3D}})
    \label{eq:min_true_profile_min_R_LS}
\end{equation}
So, if we can solve the inner 2D subproblem efficiently for every $z$, minimizing the least-squares objective reduces from a 3D search over the target coordinates to a 1D search over $z$.

\subsection{Approximate $z$-profile}
\label{subsec:approx_z_profile}
The $z$-profiled objective evaluated at a fixed $z$ in \eqref{eq:z_profile_objective} requires globally solving the inner 2D subproblem \eqref{eq:true_inner_problem} for the fixed $z$. We can also alternatively use the inner-solver constructions in Sections~\ref{section_GTRS_formulation},\ref{section_USR_formulation} and \ref{section_SDR_formulation}. we define an \emph{approximate} $z$-profile by evaluating the original R-LS objective for a candidate $z$ at a point whose horizontal coordinates are provided by an approximate inner solution. 
\begin{equation}
    J_{\text{prof.}}^{\text{approx.}}(z)
    \triangleq
    J_{\text{R-LS}}\!\left(
    \begin{bmatrix}
        \bm{X}_{\text{2D}}^{\text{approx.}}(z)\\
        z
    \end{bmatrix}
    \right),
    \label{eq:approx_z_profile}
\end{equation}
where $\bm{X}_{\text{2D}}^{\text{approx.}}(z)$ is an estimate of the global minimizer of \eqref{eq:true_inner_problem} at height $z$, and may be instantiated as
$\bm{X}_{\text{2D}}^{\text{GTRS}}(z)$, $\bm{X}_{\text{2D}}^{\text{USR}}(z)$, or $\bm{X}_{\text{2D}}^{\text{SDR}}(z)$. These approximate inner solutions have to be obtained for each candidate height $z$, using the reduced 2D formulations in section~\ref{section:SR_LS_formulations} and section~\ref{section:R_LS_formulation}. Here the ``virtual-plane'' quantities depend on $z$. In particular,
\begin{equation}
r_{\perp,k}(z)\triangleq \sqrt{y_k^2+(z_k-z)^2},\quad
\bm{A}_k^{\text{virt}}(z)\triangleq \begin{bmatrix}x_k\\-r_{\perp,k}(z)\end{bmatrix},
\label{eq:z_dependence_rperp_Avirt}
\end{equation}
and consequently the lifted affine coefficients $q_k(z)$ and $c_k(z)$ in \eqref{eq:squared_path_length_affine_form} and the stacked forms $\bm{Q}(z)$ and $\bm{b}(z)$ in \eqref{eq:Q_b} are also functions of $z$ through $\bm{A}_k^{\text{virt}}(z)$.

\begin{remark}[Exactness of approximate inner solvers]
\label{remark:exactness_inner}
Fix $z$ and let $\bm{X}_{\text{2D}}^\star(z)$ denote a global minimizer of the true inner problem \eqref{eq:true_inner_problem}. Also let $\bm{X}_{\text{2D}}^{\text{GTRS}}(z)$ denote the (globally optimal) SR-LS minimizer returned by the GTRS solver in Section~\ref{section_GTRS_formulation} using $\bm{\tilde W}$.
The statements below give \emph{sufficient} conditions for (i) exactness with respect to the \emph{surrogate} inner problem and (ii) coincidence with $\bm{X}_{\text{2D}}^\star(z)$.

\begin{enumerate}
\item \textbf{(SR-LS/GTRS agreement with the true inner problem at high SNR)} 
Under \emph{Approximation~1} (neglecting $n_k^2$ in the squared-range model) and \emph{Approximation~2} (treating $\mathrm{Var}(\epsilon_k)$ as independent of $\bm{X}_{\text{2D}}$), SR-LS and R-LS share the same \emph{noiseless} minimizer whenever it is identifiable (e.g., unique in the feasible set). Consequently, under standard regularity assumptions for weighted least-squares estimators,
\begin{equation}
\left\|\bm{X}_{\text{2D}}^{\text{GTRS}}(z)-\bm{X}_{\text{2D}}^\star(z)\right\|_2 \;\to\; 0
\qquad \text{as}\quad \mathrm{SNR}\to\infty,
\label{eq:gtrs_highsnr_agreement}
\end{equation}
(with convergence to the set of minimizers when the minimizer is non-unique).

\item \textbf{(USR exactness for SR-LS via constraint satisfaction)} 
Let $\bm{u}^{\text{USR}}(z)$ be the unconstrained WLLS minimizer in Section~\ref{section_USR_formulation} (using $\bm{\tilde W}$), and let $\bm{X}_{\text{2D}}^{\text{USR}}(z)$ be the corresponding 2D estimate.
If $\bm{u}^{\text{USR}}(z)$ satisfies the quadratic consistency constraint of the lifted SR-LS formulation, i.e.,
\begin{equation}
g\!\left(\bm{u}^{\text{USR}}(z)\right)=0,
\label{eq:usr_feasible_gtrs}
\end{equation}
then $\bm{u}^{\text{USR}}(z)$ is feasible for the GTRS and therefore
\begin{equation}
\bm{X}_{\text{2D}}^{\text{USR}}(z)=\bm{X}_{\text{2D}}^{\text{GTRS}}(z).
\label{eq:usr_tightness}
\end{equation}
In particular, whenever \eqref{eq:gtrs_highsnr_agreement} holds, the same agreement with $\bm{X}_{\text{2D}}^\star(z)$ follows for $\bm{X}_{\text{2D}}^{\text{USR}}(z)$.

\item \textbf{(SDR exactness via tight relaxation)} The semidefinite relaxation is exact when the optimal lifted matrix $\bm{Z}^\star$ is rank one. In this case, $\bm{Z}^\star$ can be written as $\bm{Z}^\star=\bar{\bm{v}}^\star (\bar{\bm{v}}^\star)^T$ for some feasible vector $\bar{\bm{v}}^\star$, and the recovered $\bm{X}_{\text{2D}}$ is a global minimizer of the original fixed-$z$ R-LS problem. In practice, tightness is indicated by $\bm{Z}^\star$ being nearly rank one. This improved tightness arises because the proposed formulation embeds both the geometric and range variables within a single PSD lift.

% Let $(\bm{V}^*,\bm{T}^*,\bm{t}^*,\bm{s}^*)$ be an optimal solution of the SDP in Section~\ref{section:R_LS_formulation}, and let $\bm{X}_{\text{2D}}^{\text{SDR}}$ denote the recovered 2D estimate obtained from $\bm{V}^*$.

% The SDR relaxation is \emph{exact} (i.e., has zero relaxation gap) if the relaxation is tight at the optimum in the following sense:

% \begin{enumerate}
%     \item \emph{Active range LMIs:} for all $k$,
%     \begin{equation}
%         (t_k^*)^2 = s_k^* 
%         = \mathrm{tr}(\bm{D}_k \bm{V}^*),
%     \end{equation}
%     i.e., the inequality $(t_k)^2 \le s_k$ (imposed via the $2\times2$ LMI in \eqref{eq:LMI_s_k_t_k}) holds with equality.

%     \item \emph{Rank-one lifting:}
%    We have
%     \begin{equation}
%         \mathrm{rank}(\bm{V}^*) = 1,
%         \qquad
%         \mathrm{rank}(\bm{T}^*) = 1.
%     \end{equation}
% \end{enumerate}

Under these conditions, the lifted matrices correspond to genuine geometric vectors and all relaxed inequalities are tight. Consequently, the SDP solution coincides with the original nonconvex R-LS solution, i.e.,
\[
\bm{X}_{\text{2D}}^{\text{SDR}} = \bm{X}_{\text{2D}}^\star.
\]
\end{enumerate}
\end{remark}

\subsection{Obtaining the final 3D estimate}
\label{section_obtaining_the_final_3D_estimate}
We can numerically obtain the R-LS objective defined in \eqref{eq:approx_z_profile} as a single argument function over the feasible interval of $z\in[z_{\min},z_{\max}]$. However, neither the true z-profile $J_{\text{prof.}}(z)$ nor the approximate z-profile $J_{\text{prof.}}^{\text{approx.}}(z)$ is guaranteed to be monotone on $[z_{\min},z_{\max}]$; thus, to minimize these objectives we need to avoid 1-D search methods that rely on unimodality. Further, since $X_{\text{2D}}^{\text{approx.}}(z)$ only approximately solves the inner problem, minimizing $J_{\text{prof.}}^{\text{approx.}}(z)$ does not guarantee global optimality of the original R-LS objective.
We therefore adopt a compact \emph{sample--polish--select} procedure.
First, uniformly sample heights
\begin{equation}
z_i = z_{\min} + \frac{i-1}{N-1}\big(z_{\max}-z_{\min}\big),\quad i=1,\dots,N,
\label{eq:z-uniform-grid}
\end{equation}
and form 3-D initializers
\begin{equation}
\widehat{X}_{\text{3D}}^{\langle0\rangle}(z_i)\triangleq
\begin{bmatrix}
X_{\text{2D}}^{\text{approx.}}(z_i)\\
z_i
\end{bmatrix}.
\label{eq:3d-init-from-profile}
\end{equation}
The approximate inner solution $X_{\text{2D}}^{\text{approx.}}(z_i)$ is obtained by choosing \emph{one} of the inner solvers in Sec.~III:
\begin{equation}
X_{\text{2D}}^{\text{approx.}}(z_i)\in
\left\{
X_{\text{2D}}^{\text{GTRS}}(z_i),\;
X_{\text{2D}}^{\text{USR}}(z_i),\;
X_{\text{2D}}^{\text{SDR}}(z_i)
\right\},
\label{eq:approx-inner-choice}
\end{equation}
corresponding to section~\ref{section_GTRS_formulation} (GTRS), section~\ref{section_USR_formulation} (USR), and section~\ref{section:R_LS_formulation} (SDR), respectively.
 
Each initializer is then refined by running $T$ Gauss--Newton iterations on the 3-D objective $J_{\text{R-LS}}$, using the GN update derived in appendix~\ref{section:GN_appendix}. Finally, we output the polished candidate with the smallest achieved residual
\begin{align}
\widehat X_{\text{3D}}
&\triangleq
\arg\min_{i\in\{1,\dots,N\}}
J_{\text{R-LS}}\!\big(X_{\text{3D}}^{(T)}(z_i)\big),
\label{eq:select-best-polished}
\end{align}
The complete algorithm is shown in algorithm~\ref{alg:zprofile-gn}.
In the numerical section, we evaluate this pipeline by instantiating $X_{\text{2D}}^{\text{approx.}}(\cdot)$ one at a time as
$X_{\text{2D}}^{\text{GTRS}}(\cdot)$, $X_{\text{2D}}^{\text{USR}}(\cdot)$, and $X_{\text{2D}}^{\text{SDR}}(\cdot)$.

\begin{algorithm}[t]
\caption{Sample--polish--select 3D estimator using a $z$-profile inner solver}
\label{alg:zprofile-gn}
\begin{algorithmic}[1]
\Require $z_{\min},z_{\max}$; number of $z$-seeds $N_z$; GN iterations $T$; data defining $J_{\text{R-LS}}(\cdot)$.
\Require Choose one inner solver for $X_{\text{2D}}^{\text{approx.}}(\cdot)$:
\begin{itemize}
\item $X_{\text{2D}}^{\text{approx.}}(\cdot)\equiv X_{\text{2D}}^{\text{GTRS}}(\cdot)$ (Section~\ref{section_GTRS_formulation}),
\item $X_{\text{2D}}^{\text{approx.}}(\cdot)\equiv X_{\text{2D}}^{\text{USR}}(\cdot)$ (Section~\ref{section_USR_formulation}),
\item $X_{\text{2D}}^{\text{approx.}}(\cdot)\equiv X_{\text{2D}}^{\text{SDR}}(\cdot)$ (Section~\ref{section:R_LS_formulation}).
\end{itemize}
\Ensure $\widehat X_{\text{3D}}$

\State $J_{\text{best}}\gets +\infty$, \quad $\widehat X_{\text{3D}}\gets \emptyset$

\For{$i=1,\dots,N_z$}
    \State $z_i \gets z_{\min} + \dfrac{i-1}{N_z-1}(z_{\max}-z_{\min})$
    \State $X_{\text{2D}}^{\text{approx.}}(z_i)\gets \textsc{Solve2D}(z_i)$ \Comment{Sec.~III}
    \State $\widehat{X}_{\text{3D}}^{\langle0\rangle} \gets 
    \begin{bmatrix}
    X_{\text{2D}}^{\text{approx.}}(z_i)\\ 
    z_i
    \end{bmatrix}$

    \For{$t=0,\dots,T-1$}
        \State $\widehat{X}_{\text{3D}}^{\langle t+1\rangle} 
        \gets 
        \textsc{GN}\!\left(\widehat{X}_{\text{3D}}^{\langle t\rangle}\right)$ 
        \Comment{Appendix~\ref{section:GN_appendix}}
    \EndFor

    \State $J \gets J_{\text{R-LS}}\!\left(\widehat{X}_{\text{3D}}^{\langle T\rangle}\right)$

    \If{$J < J_{\text{best}}$}
        \State $J_{\text{best}}\gets J$, \quad 
        $\widehat X_{\text{3D}}\gets \widehat{X}_{\text{3D}}^{\langle T \rangle}$
    \EndIf
\EndFor

\State \Return $\widehat X_{\text{3D}}$
\end{algorithmic}
\end{algorithm}

\section{Numerical Simulations and Discussion}
To evaluate the performance of the proposed Localization algorithms, we develop a Monte Carlo simulation framework in this section. As a performance benchmark, we compare the proposed estimators against the Cramér--Rao lower bound (CRLB) on the root-mean-squared error, referred to as the location error bound (PEB) \cite{shenetal}. The PEB is derived for both the 2D and 3D localization settings in Appendix~\ref{appendix:CRLB_2D_3D}.

\subsection{Simulation Setup}
Our simulation setup considers a cuboidal building of dimensions \(30\,\mathrm{m}\!\times\!30\,\mathrm{m}\!\times\!30\,\mathrm{m}\). A total of \(64\) target locations are generated by uniformly sampling the interior volume of the building. In addition, \(4096\) anchor configurations are generated by randomly sampling anchor locations from a cuboidal region of similar dimensions located outside the building. For each anchor configuration, performance is evaluated across all target locations over an SNR range of \(0\) to \(30\) dB. At each SNR value, \(100\) independent noise realizations are generated, and the average RMSE is computed over these realizations. The other simulation parameters are listed in Table \ref{tab_system_params}. The SDP algorithm was solved using the solver MOSEK \cite{mosek2026matlab}. 
\begin{table}[t]
\centering
\caption{System and Geometric Simulation Parameters}
\scriptsize
\begin{tabular}{lcl}
\toprule
\textbf{Parameter} & \textbf{Symbol} & \textbf{Value} \\
\midrule
\multicolumn{3}{l}{\textit{System and Algorithm Parameters}} \\[2pt]
Anchor Sets &$N_{\text{anchors}}$&$4096$\\
Noise Realizations &$N_{\text{noise}}$&$100$\\
Bandwidth & $B$ & $100~\text{MHz}$ \\
SNR & $SNR$ & $0-30~\text{dB}$ \\
Number of Z-seeds&$N_z$ &3,8,30\\
Number of Anchors&$K$ &5,6,7\\
Polish Steps&$T_{\text{GN}}$& 5\\
\multicolumn{3}{l}{\textit{Geometric Parameters}} \\[2pt]
Building parameter &L& $15$m\\
Building Height & $L_z$ & $2L~\text{m}$ \\
Floor Length & $L_x$ & $2L~\text{m}$ \\
Floor Width & $L_y$ & $2L~\text{m}$ \\
Target locations & $N_{\text{targets}}$ & 64 uniformly sampled targets \\
\bottomrule
\end{tabular}
\label{tab_system_params}
\end{table}

\subsection{2D Localization discussion}
For 2D localization, the target height \(z\) is assumed known, and only the horizontal coordinates \((x,y)\) are estimated from the range measurements. We consider two fundamentally different solution philosophies. The first replaces the original nonconvex R-LS problem with the surrogate SR-LS objective, for which the GTRS formulation yields an exact solution and the USR formulation provides an approximate closed-form solution. The second retains the exact R-LS objective and instead attains tractability by relaxing the nonconvex constraints, leading to a convex semidefinite programming (SDP) formulation. This naturally raises the question of whether it is more advantageous to approximate the objective while preserving exact solvability, or to retain the exact objective and rely on convex relaxation of the feasible set. \emph{A priori}, it is not clear which approach should provide better localization accuracy.

\begin{figure}[!htbp]
    \centering
\includegraphics[width=0.6\linewidth, trim=0mm 0mm 0mm 0mm, clip]{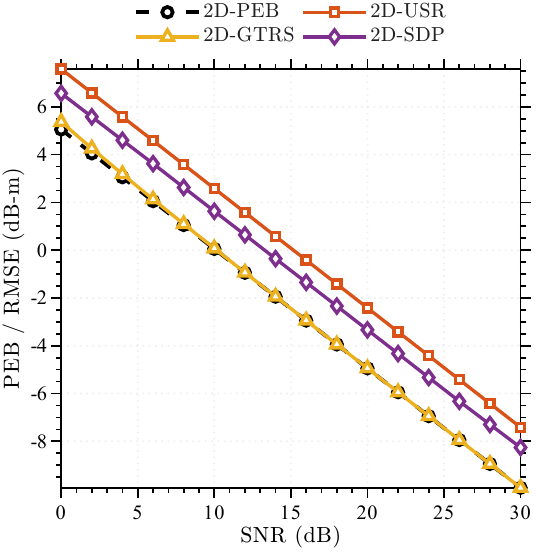}
  \caption{Monte Carlo simulation results for 2D localization with known target height \(z\), showing the average 2D-RMSE/2D-PEB vs SNR over anchor sets with \(K=6\) anchors, \(64\) indoor target locations, and \(100\) noise realizations. (a) The 2D-GTRS estimator asymptotically approaches the 2D-PEB for \(\mathrm{SNR}>5\,\mathrm{dB}\). (b) The 2D-SDP estimator provides the next best performance, while (c) The 2D-USR formulation performs the worst.}
  \label{fig_2D localization_performance}
\end{figure}

To address this question, \figref{fig_2D localization_performance} presents the 2D-PEB and 2D-RMSE versus SNR for three estimators: (a) 2D-GTRS, (b) 2D-USR, and (c) 2D-SDP. The results show that the 2D-GTRS estimator approaches the Cram\'er--Rao lower bound for \(\mathrm{SNR}>5\,\mathrm{dB}\), while the 2D-SDP estimator achieves the second-best performance. Among the considered methods, the USR-based estimator performs the worst.

The behavior of the 2D-GTRS and 2D-USR estimators can be understood by noting that both are derived from the SR-LS surrogate, which approaches the maximum-likelihood R-LS objective in the high-SNR regime, as discussed in Section~\ref{section:SR_LS_formulations}. Since the GTRS method solves this surrogate exactly, its solution asymptotically approaches that of the ML objective, thereby enabling the estimator to attain the PEB. In contrast, the USR method is obtained by relaxing the constraint set in the GTRS formulation of \eqref{eq:2D_SR_LS_GTRS}. This relaxation sacrifices exactness with respect to the surrogate problem, and consequently, the USR estimator does not attain the PEB, even though the SR-LS surrogate itself becomes asymptotically equivalent to the maximum-likelihood R-LS objective.

The 2D-SDP estimator exhibits intermediate behavior between 2D-USR and 2D-GTRS. In this case, the objective remains the exact maximum-likelihood objective, while the approximation arises from relaxing the rank-one constraint on the Gram matrix of the augmented variable, as discussed in Remark~\ref{remark:exactness_inner}.
 
\subsection{3D Localization Discussion}
\label{section_3D_localization_discussion}
\begin{figure*}[!htbp]
    \centering
    \begin{subfigure}[t]{0.35\textwidth}
        \centering
        \includegraphics[width=\linewidth,trim={0cm 0cm 0cm 0cm},clip]{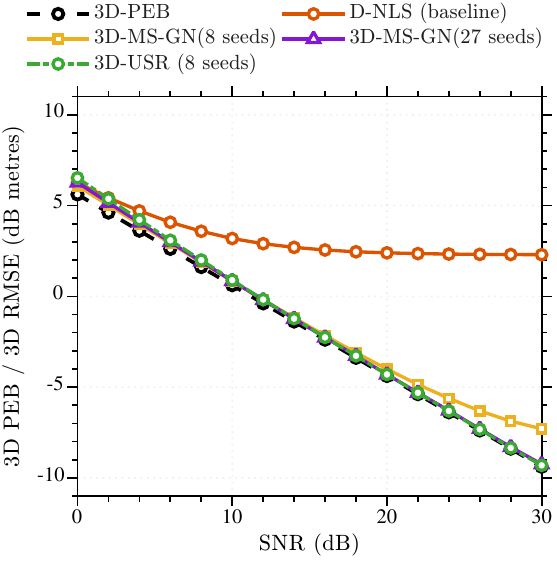}
        \caption{Multi-Start Gauss Newton vs 3D-USR}
        \label{fig_3D_usr}
    \end{subfigure}
    \hspace{0.17\textwidth}
    \begin{subfigure}[t]{0.35\textwidth}
        \centering
        \includegraphics[width=\linewidth]{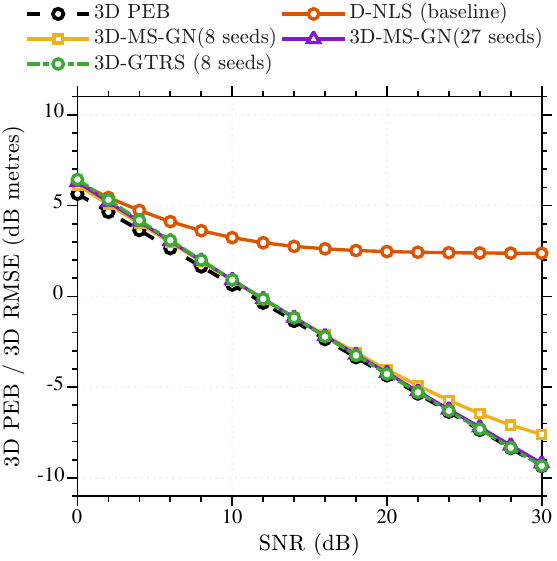}
        \caption{Multi-Start Gauss Newton vs 3D-GTRS}
        \label{fig_3D_gtrs}
    \end{subfigure}
    \caption{3D Localization Results: The single-start D-NLS baseline \cite{duggal2025diffractionaidedwirelesspositioning} does not attain the CRLB because its single arbitrary 3D initialization can lead to convergence to suboptimal local minima. The proposed sample--polish--select procedures, 3D-USR and 3D-GTRS, both approach the CRLB using only \(8\) seeds along the height dimension. By comparison, 3D-MS-GN with the same total seed count (\(2^3=8\) seeds in 3D) fails to attain the CRLB and requires \(3^3=27\) seeds to achieve similar performance. While 3D-USR and 3D-GTRS yield nearly identical accuracy, 3D-USR is computationally cheaper because USR is a one-shot estimator.}
    \label{fig_3D_results}
\end{figure*}

In this section, we evaluate the localization performance of the proposed \emph{sample-polish-select} method from Section~\ref{section_obtaining_the_final_3D_estimate} against the CRLB, D-NLS from our prior work, and multi-start Gauss--Newton (3D-MS-GN). As shown in \figref{fig_3D_results}, D-NLS does not attain the CRLB because it frequently converges to a local minimum, which motivates the use of multi-start strategies.

For 3D-MS-GN, we use \(2\) and \(3\) seeds per dimension, corresponding to \(S=8\) and \(S=27\) total seeds over the search volume. As seen in \figref{fig_3D_results}, \(S=8\) improves over D-NLS but still deviates from the CRLB, particularly at high SNR, whereas \(S=27\) yields near-CRLB performance. We compare these benchmarks with two \emph{sample-polish-select} variants based on solving the 2D subproblem using the 2D-USR and 2D-GTRS methods. The SDR-based variant is not included because of its substantially higher computational cost; see Section~\ref{section_2d_complexity}. The proposed {\em sample-polish-select} procedures are named 3D-USR in \figref{fig_3D_usr} and 3D-GTRS in \figref{fig_3D_gtrs} and both achieve near-CRLB performance with only $8$ seeds i.e. more computational efficiency than 3D-MS-GN. Also refer to Section~\ref{section_2d_complexity} for a more detailed complexity discussion.

\begin{figure}[!htbp]
    \centering
\includegraphics[width=0.6\linewidth, trim=0mm 0mm 0mm 0mm, clip]{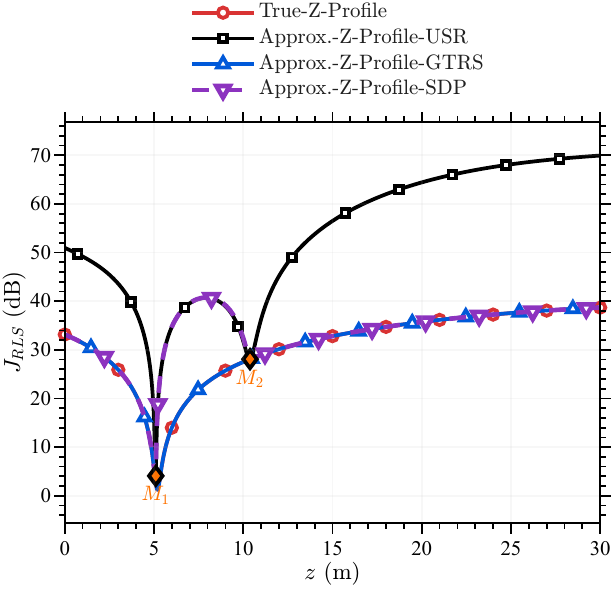}
  \caption{Maximum-likelihood R-LS objective as a function of the target height \(z\), referred to as the \(z\)-profile. The approximate \(z\)-profiles are obtained by replacing the underlying 2D solve with the corresponding GTRS-, USR-, and SDP-based solutions. At high SNR (\(25\) dB), the GTRS solution is exact for the surrogate SR-LS objective, which closely matches the R-LS objective, and therefore closely tracks the true \(z\)-profile. By contrast, the USR- and SDP-based profiles show noticeable deviation from the true profile. Two local minima are observed in the approximate z-profile: \(M_1\) at \(z=5.01\,\mathrm{m}\) and \(M_2\) at \(z=10.02\,\mathrm{m}\), with \(M_1\) being the global minimum. The final 3D estimate is obtained by searching over \(z\) to identify the global minimum and then combining the minimizing height with the corresponding solution of the associated 2D subproblem, implemented via a \emph{sample-polish-select} procedure.}
  \label{fig_z_profile}
\end{figure}

We now examine the structure of the proposed \emph{sample--polish--select} procedure through the \(z\)-profile shown in \figref{fig_z_profile}. For each fixed height \(z\), the true \(z\)-profile is obtained by solving the associated 2D subproblem using 2D-GN initialized at the true horizontal target coordinates, and then evaluating the original R-LS objective at the resulting 3D point. The approximate \(z\)-profiles are constructed in the same way, except that the fixed-\(z\) 2D subproblem is solved using the corresponding 2D-GTRS, 2D-USR, or 2D-SDP method before substituting the resulting horizontal estimate into the original R-LS objective. This distinction is important: although GTRS, USR, and SDR arise from surrogate or relaxed formulations of the fixed-\(z\) subproblem, our goal in 3D remains to minimize the original R-LS objective. For this reason, all profiles in \figref{fig_z_profile} are formed by evaluating the exact R-LS objective after obtaining a horizontal estimate for a given \(z\), rather than by plotting any surrogate objective directly. The resulting true profile represents the best achievable 3D residual as a function of height, and our goal is to identify its global minimizer, which in \figref{fig_z_profile} occurs at \(M_1\). 

A natural alternative architecture would therefore be the following: start from an initial height seed, perform a local 1D update on \(z\) alone, and then reconstruct the final 3D point by solving the fixed-\(z\) 2D subproblem using 2D-GTRS, 2D-USR, or 2D-SDP. However, even if such a procedure were able to converge to the correct height minimizer \(M_1\), the resulting 3D estimate would still generally be of the form \(\big[\bm X_{2D}^{\mathrm{approx}}(M_1)^{\top},\, M_1\big]^{\top}\), and hence would inherit the approximation error of the chosen 2D inner solver. In other words, recovering the correct minimizing height does not by itself guarantee recovery of the true 3D minimizer of the original R-LS objective, because the horizontal coordinates are still generated by an approximate fixed-\(z\) solve. Thus, an architecture based only on local search in \(z\) followed by 2D reconstruction cannot, in general, eliminate the approximation error introduced by the inner solver. 

This is precisely why the proposed method uses a \emph{sample--polish--select} structure instead. The sampled \(z\)-values are used only to generate promising 3D initializers through the fixed-\(z\) 2D subproblem. The final refinement is then performed by Gauss--Newton directly in the full 3D space on the original R-LS objective. This 3D polish step is essential for two reasons. First, the fixed-\(z\) inner solutions are approximate, since they are generated by surrogate-based or relaxed 2D solvers. Second, the height search is discretized through a finite number of seeds, so none of the sampled \(z\)-values need lie exactly at the minimizing height. A final 3D polish therefore corrects both sources of error simultaneously: it refines the horizontal coordinates beyond the approximation of the 2D inner solver and also adjusts the height away from the sampled grid, thereby moving the estimate toward the true 3D minimizer of the original R-LS objective. The select step then keeps the polished candidate with the smallest achieved R-LS value. 

This interpretation also clarifies the subtle observation that 3D-USR performs essentially the same as 3D-GTRS, even though the GTRS-based approximate \(z\)-profile almost perfectly overlaps with the true profile while the USR-based profile is visibly less accurate. At first glance, the superior profile fidelity of GTRS might suggest that 3D-GTRS should achieve markedly better final performance. However, within the proposed architecture, the inner solver need not reproduce the entire true profile accurately; it only needs to generate at least one sufficiently good 3D initializer whose basin of attraction contains the desired 3D minimizer. Once such a seed is available, the final 3D polish on the original R-LS objective can remove much of the difference between the two inner solvers. Thus, global profile fidelity is helpful, but not strictly necessary: what ultimately matters is whether the approximate fixed-\(z\) solver produces a basin-capturing initializer. This also reveals an important computational tradeoff. While GTRS provides a more faithful approximation of the true profile, USR is a one-shot estimator obtained from a single weighted least-squares solve, and therefore offers a substantial computational advantage. As a result, despite being a weaker fixed-\(z\) solver, 2D-USR can still lead to essentially the same final 3D performance as 3D-GTRS after polishing and selection, but at significantly lower inner-solver cost.

This viewpoint also highlights the role of \(N_z\). Increasing \(N_z\) reduces the spacing \(\Delta z\) between adjacent height samples and provides finer coverage of the feasible interval, thereby increasing the probability that at least one sampled height will generate a promising 3D initializer. This is a key advantage of the proposed strategy over direct multi-start search in the full 3D space: for a fixed seed budget, placing seeds over three dimensions makes them much sparser, whereas restricting the search to the single height dimension allows the same number of seeds to be placed much more densely. In this sense, the proposed method replaces a multi-start exploration over the full 3D volume by a structured and denser global exploration over only one dimension. On the other hand, if \(N_z\) is too small, important valleys of the \(z\)-profile may be missed, causing all subsequent polishes to converge to suboptimal local minima. Hence, \(N_z\) governs a robustness--complexity tradeoff: larger \(N_z\) improves coverage of the height domain, but also increases the number of fixed-\(z\) 2D solves and 3D polish operations required.

\subsection{Complexity Analysis}
Next, we examine the computational complexity of the proposed localization algorithms. Since the 2D and 3D methods have different computational structures, we treat them separately. For the 2D algorithms, complexity is evaluated as a function of the number of anchors \(K\), since they solve the reduced horizontal localization problem directly. For the 3D algorithms, complexity is instead assessed relative to MS-GN, as both methods repeatedly solve \(3\times 3\) linear systems, albeit in different contexts. In particular, the proposed 3D method follows a sample--refine--select procedure, in which candidate height values are sampled, refined through the corresponding 2D subproblem solves and a final 3D Gauss--Newton step, and the estimate with the smallest objective value is selected. Our results show that, on a per-seed basis, the sample--refine--select approach may appear to incur slightly higher cost because of the additional 2D subproblem solve. However, it requires fewer polishing steps and substantially fewer seeds than MS-GN, making it overall more computationally efficient.
\subsubsection{2D Algorithm complexity}
\label{section_2d_complexity}
For the SR-LS-based formulations, the optimization variable is the lifted vector \(\bm u\in\mathbb R^3\), whose dimension is fixed and independent of the number of anchors \(K\). In the GTRS formulation, the \(K\) measurements enter only through the aggregated quantities \(\bm M\), \(\bm m\), and \(\gamma\) in \eqref{eq:defn_M_m_delta}. Forming these quantities requires a single pass over the anchors and therefore scales linearly with \(K\). The resulting GTRS is then solved through a one-dimensional search over the Lagrange multiplier \(\lambda\). For each trial value of \(\lambda\), a \(3\times 3\) linear system must be solved; however, since this system has fixed size, its cost is essentially independent of \(K\) and in practice adds only a very small overhead. Hence, the overall complexity of 2D-GTRS scales essentially linearly with \(K\).

The same conclusion holds for 2D-USR. Its estimate is obtained from the weighted least-squares system in \eqref{eq:Q_b},
\[
\left(\mathbf Q^T\widetilde{\mathbf W}\mathbf Q\right)^{-1}\mathbf Q^T\widetilde{\mathbf W}\bm b,
\]
so its dominant cost is again the linear-in-\(K\) formation of \(\mathbf Q^T\widetilde{\mathbf W}\mathbf Q\) and \(\mathbf Q^T\widetilde{\mathbf W}\bm b\). 

By contrast, the SDP-based formulation is posed in the lifted matrix variable
\(\mathbf Z=\bar{\bm v}\bar{\bm v}^T\in\mathbb S_+^{K+3}\), where
\(\bar{\bm v}\in\mathbb R^{K+3}\). Hence, the PSD block has size
\((K+3)\times (K+3)\), and the symmetric matrix \(\mathbf Z\) contains
\(\frac{(K+3)(K+4)}{2}=\Theta(K^2)\) scalar variables. The affine constraint set also grows with \(K\): in addition to the \(K\) distance-coupling equalities in~~\eqref{eq_sdp_contraint_1}, the \(K\) nonnegativity constraints, and the normalization constraint, the triangle-inequality tightenings in~\eqref{eq_sdp_contraint_2} and~\eqref{eq_sdp_contraint_3} introduce \(\Theta(K^2)\) pairwise constraints. Therefore, unlike GTRS and USR, the SDP does not admit a fixed-dimensional core problem.

When solved using a generic conic interior-point method, such as the optimizer used by MOSEK, the computational cost is polynomial in both the PSD block size and the number of affine constraints. Since here \(n=K+3=\Theta(K)\) and \(m=\Theta(K^2)\), the runtime grows superlinearly with \(K\). Under a dense generic interior-point model, this gives a representative worst-case per-iteration scaling of order \(O(K^6)\).

\subsubsection{3D Algorithm complexity}
For the 3D methods, a direct complexity comparison is less straightforward because the computational effort is allocated differently across the algorithms. In 3D-MS-GN, seeds are placed throughout the full 3D search space, and each seed is refined through iterative Gauss--Newton updates, where every iteration requires solving a \(3\times 3\) linear system. In contrast, the proposed sample--polish--select approach places seeds only along the 1D height dimension. For each seed, it first solves an associated 2D subproblem and then performs a final 3D Gauss--Newton polish, which again involves iterative solutions of a \(3\times 3\) linear system. While this can increase the cost per seed, the proposed method typically requires far fewer seeds and fewer polishing iterations because the 2D subproblem provides a more accurate initializer, often closer to the true solution. Owing to these competing effects, numerical timing results are more informative than asymptotic complexity alone. In practice, the main computational savings come from reducing the seeding dimension and decreasing the number of Gauss--Newton iterations needed in the polish step.

\begin{figure}[!htbp]
    \centering
\includegraphics[width=0.6\linewidth, trim=0mm 0mm 0mm 0mm, clip]{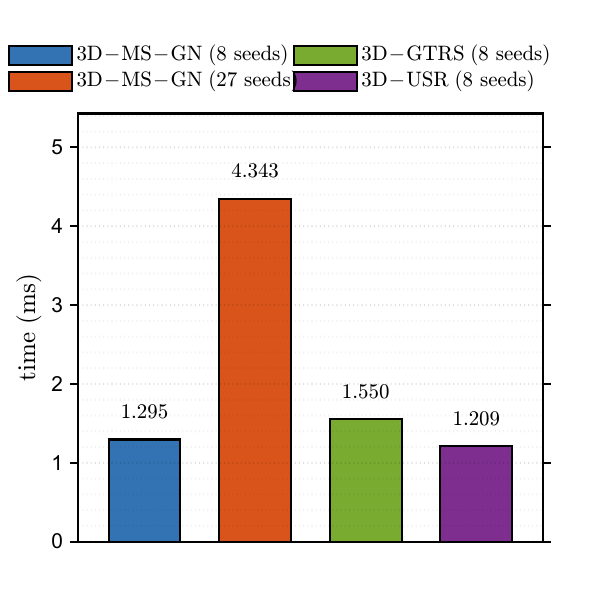}
  \caption{Computation time comparison of 3D-MS-GN, 3D-USR, and 3D-GTRS. For 3D-MS-GN, results are shown for \(8\) and \(27\) seeds generated over the full 3D search space. While the \(8\)-seed 3D-MS-GN configuration has the lowest cost among the multi-start GN baselines, it does not attain the CRLB. Increasing the number of seeds to \(27\) improves estimation performance, but at a substantially higher computational cost. In contrast, both 3D-USR and 3D-GTRS achieve near-CRLB performance while requiring significantly less computation than \(27\)-seed 3D-MS-GN, highlighting the efficiency of the proposed one-dimensional seeding strategy.}
  \label{fig_timing_3D_algos}
\end{figure}

\section{Conclusion}
This paper developed a new estimator for mixed LOS/NLOS localization based on a unified diffraction-aware path-length model that transitions smoothly between LOS and diffraction-dominated NLOS propagation, thereby avoiding explicit path labeling. A key structural result is that, for known target height, the proposed diffraction geometry admits a virtual-anchor formulation that yields an exact Euclidean embedding of the reduced 2D problem. This enables three supporting 2D estimators based on GTRS, USR, and SDR formulations. Building on this structure, the main contribution is a new 3D estimator that profiles out the horizontal coordinates and reduces the original nonconvex maximum-likelihood problem to a one-dimensional search over target height, followed by local Gauss–Newton refinement through a sample–polish–select procedure. Simulation results indicate that the proposed approach provides strong robustness to initialization and near-CRLB performance while requiring substantially less computations than conventional multistart 3D Gauss–Newton methods.
\appendix
\subsection{Gauss Newton Method}
\label{section:GN_appendix}
Let \(r_k\) denote the ranging measurement from the \(k^{\text{th}}\) anchor, and let \(p_k(\bm{X}_{3D})\) denote the corresponding measurement model in \eqref{eq_diffraction_path_length_approx} between the \(k^{\text{th}}\) anchor and the target. To obtain an estimate of the target's location $\bm{X}_{3D}$ we use the least squares criterion
\begin{align}\label{eq:RLS}
 \min_{\bm{X}_{3D}} \sum_{k=1}^{K}w_k\left(r_k-p_{k}(\bm{X}_{3D})\right)^2.
\end{align}
Here, the weights $w_k=1/\sigma_k^2$ are related to the uncertainty in the ranging measurements. Observe that the path length $p_{k}(\bm{X}_{3D})$ in the least squares objective is a nonlinear function of the target location $\bm{X}_{3D}$. Therefore, to estimate the 3D location, we can use an iterative non-linear estimation technique such as the Gauss-Newton method \cite{kay1993fundamentals}. The update step is given by
\begin{equation}  \hat{\bm{X}}_{3D}^{\langle t+1 \rangle} = \hat{\bm{X}}_{3D}^{\langle t \rangle} + \Delta^{\langle t \rangle}.
\end{equation}
Here, the superscript $t$ denotes the iteration index. The vector $\hat{\bm{X}}_{3D}^{\langle t+1 \rangle} \in \mathbb{R}^3$ represents the estimate at iteration $t+1$, and $\Delta^{\langle t \rangle} \in \mathbb{R}^3$ denotes the update step computed at iteration $t$. 
For compactness define the following terms for the $k^{th}$ anchor at iteration $t$ 
\begin{align}
    a_k& \triangleq x_k-x,\; r_{\perp,k} \triangleq  \sqrt{y_k^2+(z-z_k)^2},\nonumber \\ 
    b_k& \triangleq r_{\perp,k} + y,\;p_k=\sqrt{a_k^2+b_k^2}.
\end{align}
Hence the  Jacobian matrix $\bm{J} \in\mathbb{R}^{K\times 3}$ for the path length model at iteration $t$ can be defined as
\begin{align}    
\mathbf{J}
=
\begin{bmatrix}
-\dfrac{a_1}{p_1} & \dfrac{b_1}{p_1} & -\dfrac{b_1(z_1-z)}{p_1r_{\perp,1}}\\[10pt]
-\dfrac{a_2}{p_2} & \dfrac{b_2}{p_2} & -\dfrac{b_2(z_2-z)}{p_2r_{\perp,2}}\\
\vdots & \vdots & \vdots\\
-\dfrac{a_K}{p_K} & \dfrac{b_K}{p_K} & -\dfrac{b_K(z_K-z)}{p_Kr_{\perp,K}}
\end{bmatrix}.
\end{align}
Finally, the update step is given by

\begin{equation}
    \Delta^{\langle t \rangle} = \left (\bm{J}^{\langle t \rangle T}\bm{W} \bm{J}^{\langle t \rangle} \right)^{-1}\bm{J}^{\langle t \rangle T}\bm{W}\left(\bm{r} -\bm{p}^{\langle t\rangle}\right).  
\end{equation}
Here, $\bm{W} \triangleq \mathrm{diag}(w_1,\ldots,w_K)$.

\begin{remark}
The Gauss--Newton method is initialized from an arbitrary starting point. Under standard regularity conditions, it is only locally convergent \cite{kay1993fundamentals}, and may therefore converge to a stationary point rather than the global minimizer of the least-squares objective. As a result, the resulting 3D estimate is not generally expected to attain the Cram\'er--Rao lower bound (CRLB), i.e., the position error bound (PEB) on the RMSE \cite{shenetal}. This behavior is also seen in the simulation results for D-NLS in \figref{fig_3D_results}.
\end{remark}

\subsection{KKT conditions SR-LS GTRS}
\label{section_KKT}

Form the Lagrangian for the GTRS in \eqref{eq:2D_SR_LS_GTRS} as
\begin{equation}
\mathcal{L}(\bm{u},\lambda)=\bm{f(u)}+\lambda \bm{g(u)},
\end{equation}
where $\lambda\in\mathbb{R}$ since the constraint in \eqref{eq:2D_SR_LS_GTRS} is an equality.
\par
The KKT stationarity condition $\nabla_{\bm{u}}\mathcal{L}(\bm{u^\star},\lambda^\star)= \bm{0}
$ yields
\begin{equation}
(\mathbf{M}+\lambda\mathbf{H})\bm u=-(\bm m+\lambda\bm h).
\label{eq:app_stationarity}
\end{equation}
For any $\lambda$ such that $(\mathbf{M}+\lambda\mathbf{H})$ is invertible,
\begin{equation}
\bm{u}(\lambda)=-(\mathbf{M}+\lambda\mathbf{H})^{-1}(\bm m+\lambda\bm h).
\label{eq:app_u_of_lambda}
\end{equation}
Note this could be solved with numerically stable fast methods like Cholesky factorization avoiding the matrix inversion operation \cite{golub2013matrix}. 
Imposing feasibility by substituting this $\lambda$ into the constraint of \eqref{eq:2D_SR_LS_GTRS} by $g(\bm{u}(\lambda))$ gives the scalar equation
\begin{equation}
\phi(\lambda)\triangleq \bm u(\lambda)^T\mathbf{H}\bm u(\lambda)+2\bm h^T\bm u(\lambda)=0,
\label{eq:app_phi}
\end{equation}
so any KKT point satisfies $\phi(\lambda^\star)=0$ and $\bm u^\star=\bm u(\lambda^\star)$, reducing \eqref{eq:2D_SR_LS_GTRS} to a 1D root search in $\lambda$.

Differentiating \eqref{eq:app_stationarity} with respect to $\lambda$ gives
\begin{equation}
(\mathbf{M}+\lambda\mathbf{H})\bm u'(\lambda)=-(\mathbf{H}\bm u(\lambda)+\bm h),
\label{eq:app_uprime_lin}
\end{equation}
hence
\begin{equation}
\bm u'(\lambda)=-(\mathbf{M}+\lambda\mathbf{H})^{-1}(\mathbf{H}\bm u(\lambda)+\bm h).
\label{eq:app_uprime}
\end{equation}
Differentiating \eqref{eq:app_phi} and substituting \eqref{eq:app_uprime} yields
\begin{equation}
\phi'(\lambda)=
-2\big(\mathbf{H}\bm u(\lambda)+\bm h\big)^T
(\mathbf{M}+\lambda\mathbf{H})^{-1}
\big(\mathbf{H}\bm u(\lambda)+\bm h\big).
\label{eq:app_phi_prime}
\end{equation}
Therefore, over any interval where $(\mathbf{M}+\lambda\mathbf{H})\succ 0$, we have $\phi'(\lambda)\le 0$, i.e.,
$\phi(\lambda)$ is monotone nonincreasing. Consequently, once a bracket $[\lambda_L,\lambda_U]$ is found within this
positive-definite region with $\phi(\lambda_L)\phi(\lambda_U)\le 0$, the root $\lambda^\star$ is unique and can be
obtained reliably via bisection; the corresponding primal optimizer is then $\bm u^\star=\bm u(\lambda^\star)$ from
\eqref{eq:app_u_of_lambda}.

 \subsection{2D/3D localization CRLB}
 \label{appendix:CRLB_2D_3D}
For $d\in\{2,3\}$, let $\bm{\theta}_d\in\mathbb R^d$ denote the unknown location and
$r_k=p_k(\boldsymbol{\theta}_d)+n_k$ with $n_k\sim\mathcal N(0,\sigma_k^2)$ denote the mathematical model behind the range measurements. Define the gradient vector corresponding to the $k^{\text{th}}$ anchor location with respect to the unknown location vector as 
\[\bm g_{k,d}\triangleq \nabla_{\boldsymbol{\theta}_d}p_k(\boldsymbol{\theta}_d).\]
Then the Fisher Information Matrix (FIM) $\bm{\mathcal{I}}_d(\bm{\theta}_d) \in \mathbb{R}^{d\times d}$ corresponding to the estimation parameters $\bm{\theta}_d$ can be obtained using the standard result in Gaussian noise \cite{kay1993fundamentals}
\[ \bm{\mathcal{I}}_d(\bm{\theta}_d)=\sum_{k=1}^K \frac{1}{\sigma_k^2}\bm{g}_{k,d}\bm{g}_{k,d}^{T},
\]
and the PEB is
\[
\text{PEB}_d=\sqrt{\mathrm{tr}\!\left(\bm{\mathcal{I}}_d(\bm{\theta}_d)^{-1}\right)}.
\]
For 2D localization with known height $z_0$, $p_k(\bm{X}_{\text{2D}};z_0)=\|\bm{X}_{\text{2D}}-\tilde{\bm{a}}_k(z_0)\|_2$,
\[
\mathbf g_{k,2}
=
\frac{\bm{X}_{\text{2D}}-\tilde{\bm{a}}_k(z_0)}
{\|\bm{X}_{\text{2D}}-\tilde{\bm{a}}_k(z_0)\|_2},
\]
where $\tilde{\bm{a}}_k(z_0)=[x_k,\,-r_{\perp,k}(z_0)]^{\top}$ and
$r_{\perp,k}(z_0)=\sqrt{y_k^2+(z_k-z_0)^2}$.
For 3D we have
\[
p_k(\bm{X}_{\text{3D}})=\sqrt{(x-x_k)^2+\bigl(y+r_{\perp,k}(z)\bigr)^2},
\]
with $r_{\perp,k}(z)=\sqrt{y_k^2+(z_k-z)^2}$, and
\[
\bm{g}_{k,3}
 =
\frac{1}{p_k(\bm{X}_{\text{3D}})}
\begin{bmatrix}
x-x_k\\
y+r_{\perp,k}(z)\\
\dfrac{\bigl(y+r_{\perp,k}(z)\bigr)(z-z_k)}{r_{\perp,k}(z)}
\end{bmatrix}.
\]

\bibliographystyle{IEEEtran}{
\footnotesize
\bibliography{refs}
}
\end{document}

%% file: notation.tex
\def\nb0{{\mathbf{0}}}
\def\nb1{{\mathbf{1}}}

\newtheorem{remark}{Remark}

\def\figref#1{Fig.\,\ref{#1}}%